\documentstyle[aps,prl,preprint,floats,epsf,epsfig]{revtex}

\begin{document}

\tighten

\title{Analyses of $D^+ \to K^0_SK^+$ and $D^+ \to K^0_S\pi^+$}  

\date{\today}
\maketitle

\vspace*{-3.cm}
\begin{flushright}
CLNS 96/1449
CLEO 96-20
\end{flushright}
\vskip 2.5cm

\begin{abstract} 
Using data collected with the CLEO II detector at the
Cornell Electron Storage Ring, we present new measurements of the
branching fractions for $D^+ \to K_S K^+$ and $D^+ \to K_S \pi^+$.
These results are combined with other CLEO measurements to extract
the ratios of isospin amplitudes and phase shifts for 
$D \to KK$ and $D \to K\pi$.
\end{abstract}

\begin{center}
M.~Bishai,$^{1}$ J.~Fast,$^{1}$ E.~Gerndt,$^{1}$
J.~W.~Hinson,$^{1}$ N.~Menon,$^{1}$ D.~H.~Miller,$^{1}$
E.~I.~Shibata,$^{1}$ I.~P.~J.~Shipsey,$^{1}$ M.~Yurko,$^{1}$
L.~Gibbons,$^{2}$
% ...
\end{center}
 
\small
\begin{center}
$^{1}${Purdue University, West Lafayette, Indiana 47907}\\
$^{2}${University of Rochester, Rochester, New York 14627}\\
% ...
$^{13}${Cornell University, Ithaca, New York 14853}\\
% ...
$^{25}${University of Oklahoma, Norman, Oklahoma 73019}
\end{center}

Strong final-state interactions (FSI) in 
nonleptonic weak decays of hadrons obscure
the underlying weak interactions.
The problem is particularly acute for 
the $D$ meson, as its mass lies in a resonance-rich 
region \cite{Sutherland,Lipkin,Dono1}.  
Elastic (i.e. $\pi\pi$ stays as $\pi\pi$) and inelastic FSI 
rotate the isospin amplitudes \cite{Kamal1}.
These isospin amplitudes may be inferred by combining
measurements of branching fractions.
This Letter reports new measurements of the 
$D^+ \to K_S K^+$ and $D^+ \to K_S \pi^+$ branching fractions.  
We combine these results with previous CLEO measurements of 
$D^0$ branching fractions \cite{kk,kzpz,kmpp} 
to obtain the first measurement of the isospin amplitudes and 
phase shift difference for $D \to KK$ and improved values of these
quantities for $D \to K \pi$.

The CLEO II detector \cite{NIM} is designed to measure 
charged particles and photons with high efficiency and precision.
This analysis is based on 3.12 $fb^{-1}$ of
data collected at the $\Upsilon(4S)$ resonance and 1.72 $fb^{-1}$ 
60 MeV below the $\Upsilon(4S)$.  Hadronic events are selected
by requiring at least three charged tracks, a total detected energy of
at least 0.15 $E_{c.m.}$, and a primary vertex within 5 cm along the
beam ($z$) axis of the interaction point. 

Candidate $K_S$ mesons are detected in the $K_S \to \pi^+ \pi^-$ mode.
They are reconstructed by combining pairs of oppositely charged 
tracks, each with an impact parameter in $r-\phi$ of
greater than four times the measurement uncertainty.  
The track pair must also pass a $\chi^2$ cut based
on the the distance in $z$ between the two tracks at their $r-\phi$ 
intersection point.
The invariant mass of the track pair must
be within 15 MeV of the known $K_S$ mass.

Charged pion and kaon candidates must pass 
minimum track-quality requirements.
To reduce combinatoric background 
in the $D^+ \to K^- \pi^+ \pi^+$ channel,
we require that the specific ionization ($dE/dx$) of the $K^+$ 
candidate be within 3 standard deviations ($\sigma$) 
of that expected for a kaon.  Tighter cuts are 
applied on the $K^+$ candidates in the $D^+ \to K_S K^+$ mode
because of a large background from $D^+ \to K_S \pi^+$ decays.
The measured $dE/dx$ must be within 2$\sigma$ of that expected 
for a kaon and
at least 0.25$\sigma$ lower than that expected for a pion.

We then reconstruct $D^+$ candidates 
from the $K_S$, $K^+$, and $\pi^+$
candidates in the signal modes 
$D^+ \to K_S K^+$ and $D^+ \to K_S \pi^+$, 
and the normalization mode $D^+ \to K^- \pi^+ \pi^+$.  
In the $D^+ \to K_S \pi^+$ mode, we observe a large background from
events in which a $K_S$ candidate is combined with a random slow pion.
We therefore require
cos$(\theta_{K_S}) < 0.8$, where $\theta_{K_S}$ is the
angle between the $K_S$ in the $D^+$ rest frame and the $D^+$ direction in
the laboratory frame.  This requirement is also imposed on the
$D^+ \to K_S K^+$ mode.

We require that every $D^+$ candidate also be a product of the 
decay $D^{*+} \to D^+ \pi^0$. 
The low-momentum $\pi^0$ provides a clean tag for the $D^+$. 
Pairs of electromagnetic showers detected by CLEO's CsI(Tl) crystal
calorimeter are combined to form $\pi^0$ candidates, 
which must have $M(\gamma\gamma)$ within 2.5$\sigma$ (about 15 MeV) of 
$m_{\pi^0}$.  Both daughter photons must be detected in 
the ``barrel'' region of the detector, have energies of greater than 30
MeV, and deposit most of their energy in a compact group of crystals
\cite{e9e25}.  

Since $D^{*+}$ fragmentation is
relatively hard \cite{PDG} and combinatoric background comes mostly from
low-momentum tracks, we impose a cut of X 
$\equiv p(D^{*+})/p(D^{*+}_{max}) > 0.55$.  For each event we
calculate $\Delta M$, the difference between the reconstructed 
$D^{*+}$ and $D^+$ masses.  We require 
$\Delta M$ to be within 2.5 MeV (3$\sigma$) of the
known mass difference.  

Events in which a random slow $\pi^0$ is combined with 
a correctly reconstructed $D^+$ will contribute to the 
peak in $M(D^+)$ \cite{dplus}, 
but will not peak in the $\Delta M$ distribution.  
In order to remove this background, 
we perform a sideband subtraction in $\Delta M$.
The resulting invariant-mass distributions for all events 
passing the cuts is shown in Figure 1.

\nopagebreak
\begin{figure}
\centering
\mbox{\psfig{figure=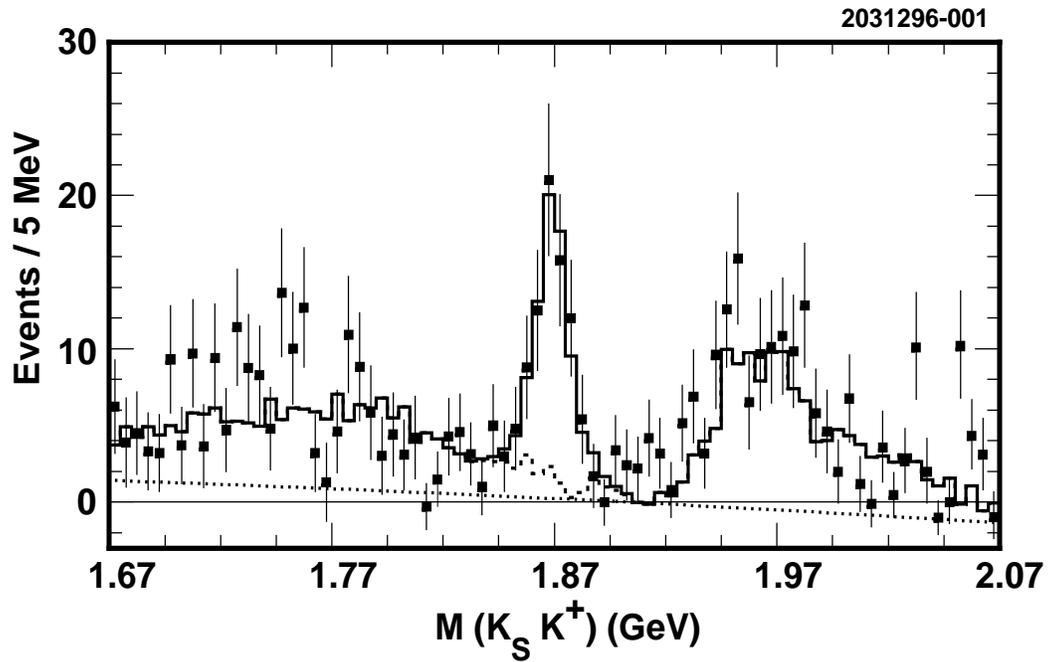,width=5.5in,height=3.5in}}
\centering
\mbox{\psfig{figure=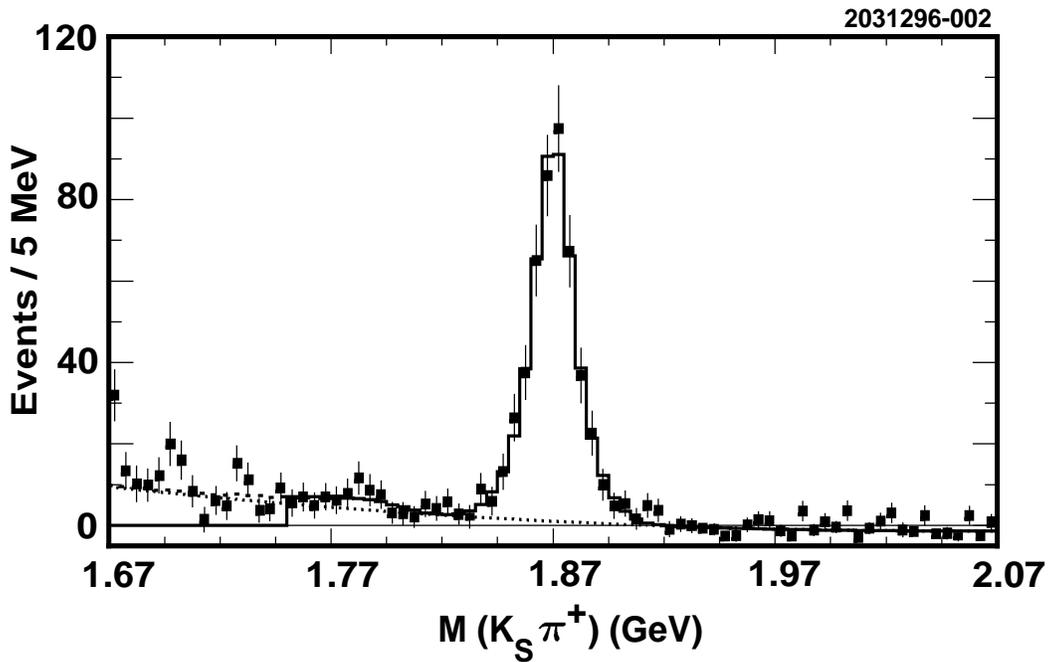,width=5.5in,height=3.5in}}
\caption{The experimental data and fits.  Data are the solid points.
The solid lines are fits to the data; 
the dashed lines are the combinatoric background components of the fit.
(a) Upper plot:  $M(K_S K^+)$.  
(b) Lower plot:  $M(K_S \pi^+)$.  
}
\end{figure}
\nopagebreak

The reconstruction efficiencies for the signal and normalization modes
were estimated using a GEANT-based Monte Carlo 
simulation \cite{GEANT} of the CLEO II detector.
Furthermore, to study the combinatoric background in 
$M(D^+)$ for each mode, we 
ran a full Monte Carlo simulation that included all particle decay
processes except for the signal mode and peaking backgrounds from
other specific decay modes.   In all three decay modes, the 
combinatoric backgrounds 
are smooth and are fit well using a quadratic polynomial.

Figure 1(a) shows the $K_S K^+$ invariant mass spectrum.
The peak at about 1.95 GeV is from $D^+ \to K_S \pi^+$ 
events in which the $\pi^+$ is misidentified as a kaon.
The broad peak  in the low mass region is from  
$D^+ \to K_S \rho^+$, $D^0 \to K_S \rho^0$, and 
$D^0 \to K^*(892)^-\pi^+$ events.  In each of these events, a 
charged pion is identified as a kaon, and the other pion is
undetected.  The shapes of these peaks were obtained from Monte Carlo.  
The relative normalization of each mode 
was fixed to PDG \cite{PDG} values, and
the overall normalization of the sum was allowed to float in the fit.
The combinatoric background is parametrized by a quadratic polynomial.
The signal is fit with a sum of two Gaussians.  The
ratios of the widths and areas of the two Gaussians are obtained from
signal Monte Carlo, and the overall width is allowed to float.
We find a signal yield of $70.3 \pm 12.1$ events at the $D^+$ mass.  
The reconstruction efficiency, $\epsilon$, is $(6.91 \pm 0.23)\%$.
As a cross-check, we obtain
$B(D^+ \to K_S K^+)/B(D^+ \to K_S \pi^+) = 0.28 \pm 0.07$
from the normalization of the reflection background component,
which is consistent with the direct measurement.

Figure 1(b) shows the $K_S \pi^+$ invariant mass spectrum.
The background below 1.75 GeV is primarily 
$D^+ \to \bar{K^0} \ell^+ \nu_\ell$,
which is small and far from the signal, so we exclude this region.
The region between 1.75 GeV and 1.80 GeV is enhanced by 
$D^+ \to \bar{K^0} K^+$;  we obtain the shape of this background with
Monte Carlo and include it in the fit, allowing the normalization to float.
The combinatoric background and signal are fitted 
using the same procedure as
above.  We observe
$473 \pm 26$ events with $\epsilon = (9.32 \pm 0.27)\%$.  

In the normalization mode of
$D^+ \to K^- \pi^+ \pi^+$ we observe $5430 \pm 108$ events with
$\epsilon = (12.43 \pm 0.19)\%$.

The systematic errors are summarized in Table I.  To study
the particle identification cuts, we use a
kinematically identified sample of kaons from the decay chain
$D^{*+} \to D^0 \pi^+$, $D^0 \to K^- \pi^+$.  
The cut efficiency as a function of kaon momentum for both data and
Monte Carlo is measured, then integrated over the
$K^+$ momentum distribution of Monte Carlo $D^+ \to K_S K^+$ events.
This yields an overall momentum-weighted efficiency.  We find 
$\epsilon_{MC}(D^+ \to K_S K^+) / \epsilon_{data}(D^+ \to K_S K^+) 
= 1.100 \pm 0.030$,
so a correction factor of 1.10 is applied to the efficiency-corrected
yield, $N$, of $D^+ \to K_S K^+$.  
From a similar study of the secondary 
vertex requirements, we obtain correction factors
of $1.030 \pm 0.014$ for $N(D^+ \to K_S \pi^+)$
and $1.036 \pm 0.011$ for $N(D^+ \to K_S K^+)$.

The $K^- \pi^+ \pi^+$ systematic error is due to differences between
the Monte Carlo simulation and data in the Dalitz plot distribution of
$D^+ \to K^- \pi^+ \pi^+$ events.  The systematic error in the fitting
procedure was estimated by varying the Monte Carlo background shapes,
fitting functions, fit regions, and bin sizes.
The systematic error for Monte Carlo tracking efficiency is small
because we measure ratios of branching fractions, and all decay modes
have a final state of three charged tracks.

\begin{table}[h]
\begin{center}
\begin{tabular}{l c c}
Systematic bias & $\frac {B(K_S K^+)} {B(K_S \pi^+)}$ 
& $\frac {B(K_S \pi^+)} {B(K^- \pi^+ \pi^+)}$ \\ \hline
$K^+$ particle ID             & $4.0\%$ & $1.0\%$ \\
$K_S$ detection efficiency    & $1.5\%$ & $2.0\%$ \\
$K^- \pi^+ \pi^+$ Dalitz structure & $0.0\%$ & $3.5\%$ \\
$\pi^0$ cuts                  & $1.2\%$ & $1.2\%$ \\
Sideband determination        & $2.7\%$ & $2.7\%$ \\
Fitting Procedure             & $6.4\%$ & $1.5\%$ \\
Signal shape                  & $2.8\%$ & $2.8\%$ \\
Tracking efficiency           & $1.0\%$ & $1.0\%$ \\ \hline
TOTAL                         & $8.7\%$ & $6.1\%$ \\
\end{tabular}
\end{center}
\label{tab:skim}
\caption[h] {Systematic error summary.  For each ratio measurement,
we list each contribution to the systematic error in terms of 
percentage of the measurement.}
\end{table}

The final results are:
\begin{displaymath}
\frac{B(D^+ \to K_S K^+)}{B(D^+ \to K_S \pi^+)} 
= 0.222 \pm 0.041 \pm 0.019
\end{displaymath}
\begin{displaymath}
\frac{B(D^+ \to K_S K^+)}{B(D^+ \to K^- \pi^+ \pi^+)} 
= 0.0386 \pm 0.0069 \pm 0.0037
\end{displaymath}
\begin{displaymath}
\frac{B(D^+ \to K_S \pi^+)}{B(D^+ \to K^- \pi^+ \pi^+)} 
= 0.174 \pm 0.012 \pm 0.011
\end{displaymath}
where the first error is statistical and the second error is systematic.

To find absolute branching fractions, 
the last two results are combined with
the PDG value $B(D^+ \to K^- \pi^+ \pi^+) = (9.1 \pm 0.6)\%$ \cite{PDG}.
When converting the measurements of $K_S$ branching fractions to 
branching fractions involving
$\bar{K}^0$, we must take into account the possibility of
interference between doubly Cabbibo suppressed
and favored modes \cite{BigYam}.  
The amplitudes interfere with a magnitude of
roughly $2\tan^2(\theta_C)\cos(\beta)$, where $\theta_C$ is the Cabbibo
angle and $\beta$ is an interference phase.
Because $\beta$ is unknown we assign an $10\%$ uncertainty 
to $B(D^+ \to \bar{K}^0 \pi^+)$.
There is no such interference in $D^+ \to \bar{K}^0 K^+$.  We obtain
\begin{displaymath}
B(D^+ \to \bar{K}^0 K^+) = (0.70 \pm 0.12 \pm 0.07 \pm 0.05)\%
\end{displaymath}
\begin{displaymath}
B(D^+ \to \bar{K}^0 \pi^+) 
= (3.17 \pm 0.21 \pm 0.19 \pm 0.21 \pm 0.32)\%
\end{displaymath}
where the third error is due to uncertainty in the normalization branching
fraction and the fourth error is due to the possibility of DCSD interference.
Our result for $B(D^+ \to \bar{K}^0 K^+) / B(D^+ \to \bar{K}^0 \pi^+)$
is $3.6\sigma$ higher than $\tan^2(\theta_C)$, consistent with the
expectation that
destructive interference suppresses the $D^+ \to \bar{K}^0 \pi^+$ 
rate \cite{lifetime}.  

The amplitudes for the three $D \to KK$ decays may be 
decomposed into isospin amplitudes:
\begin{displaymath}
A^{+-} = \frac{1}{\sqrt{2}}(A_1 + A_0)
\end{displaymath}
\begin{displaymath}
A^{00} = \frac{1}{\sqrt{2}}(A_1 - A_0)
\end{displaymath}
\begin{displaymath}
A^{+0} = \sqrt{2}A_1
\end{displaymath}
where $A^{+-} \equiv \langle K^+K^- | H | D^0 \rangle$,
$A^{00} \equiv \langle K^0 \bar{K}^0 | H |D^0 \rangle$, and
$A^{+0} \equiv \langle K^+ \bar{K}^0 | H |D^+ \rangle$.  
We have assumed that the Hamiltonian reponsible for these decays has
isospin structure $|I, I_3> = |\frac{1}{2}, +\frac{1}{2}>$.  
This is true for
all Standard Model decay processes,
except for two:  the $s\bar{s}$-popping W-exchange
diagram of $D^0 \to \bar{K}^0K^0$, and the $D^+$ annihilation diagram
of $D^+ \to \bar{K}^0 K^+$.  Since these diagrams are helicity-suppressed
and require $s\bar{s}$ popping, they are expected to be small.

From these equations one can express the ratio of isospin amplitudes
and the isospin phase angle difference in terms 
of measured branching fractions:

\begin{displaymath}
\left|\frac{A_1}{A_0}\right|^2 = 
\frac{\Gamma^{+0}}
{2\Gamma^{+-}+2\Gamma^{00}-\Gamma^{+0}}
\end{displaymath}
\begin{displaymath}
\cos(\delta_{KK}) = 
\frac{\Gamma^{+-}-\Gamma^{00}}
{\sqrt{\Gamma^{+0}}\sqrt{2\Gamma^{+-}+2\Gamma^{00}-\Gamma^{+0}}}
\end{displaymath}
for $D \to KK$, where $\delta_{KK} \equiv \arg(A_1/A_0)$.  

In the Cabbibo-favored $D \to K\pi$ system, the Hamiltonian has
isospin structure $|1, +1>$.
The isospin decomposition and the equations for 
$|A_{\frac{3}{2}}/A_{\frac{1}{2}}|$ and 
$\delta_{K\pi} \equiv \arg(A_{\frac{3}{2}}/A_{\frac{1}{2}})$ 
in the $D \to K\pi$ system are also similar to those of 
$D \to KK$ and may be found elsewhere \cite{BSW}.

CLEO has now measured the six branching fractions necessary to 
calculate the amplitude ratios and phase shifts in $D \to KK$ and 
$D \to K\pi$ \cite{kk,kzpz,kmpp}.
All branching fractions are written in terms of a fraction of
{\sf B} $\equiv B(D^0 \to K^- \pi^+)$, in order to avoid additional
statistical error from the uncertainty in {\sf B}.
We use the CLEO result 
$B(D^+ \to K^- \pi^+ \pi^+) = (2.35 \pm 0.16 \pm 0.16)${\sf B} \cite{vj} 
and the PDG fit result
$B(D^0 \to \bar{K}^0 \pi^+ \pi^-) = (1.41 \pm 0.11)${\sf B} \cite{PDG}.
The results are listed in Table II.

\begin{table}[h]
\begin{center}
\begin{tabular}{c|c c}
Measurement & $D \to KK$ & $D \to K\pi$ \\ \hline
$\frac{\displaystyle \Gamma^{+-}}{\displaystyle \Gamma_{D^+}}$ 
& $(0.116 \pm 0.010)${\sf B} & {\sf B} \\
$\frac{\displaystyle \Gamma^{00}}{\displaystyle \Gamma_{D^0}}$ 
& $(0.014 \pm 0.004)${\sf B} & $(0.620 \pm 0.126)${\sf B} \\
$\frac{\displaystyle \Gamma^{+0}}{\displaystyle \Gamma_{D^0}}$ 
& $(0.182 \pm 0.041)${\sf B} & $(0.819 \pm 0.136)${\sf B} \\ \hline
amplitude ratio & $\left|\frac{A_1}{A_0}\right| 
= 0.61 \pm^{0.11}_{0.10}$ & 
$\left|\frac{A_{\frac{3}{2}}}{A_{\frac{1}{2}}}\right| = 0.27 \pm 0.03$ \\
$\cos \delta$ & $0.88 \pm^{0.10}_{0.08}$ & $-0.12 \pm^{0.23}_{0.21}$ \\
\end{tabular}
\end{center}
\label{tab:skim}
\caption[h] {Isospin analysis inputs and results.  
$\Gamma^{+-} \equiv |A^{+-}|^2$, 
$\Gamma^{00} \equiv |A^{00}|^2$, 
$\Gamma^{+0} \equiv |A^{+0}|^2$, and
{\sf B} $\equiv B(D^0 \to K^- \pi^+)$.   }
\end{table}

In conclusion, we find that the isospin phase shift difference in $D \to
KK$ is significantly smaller than that of both $D \to K\pi$ and $D \to
\pi\pi$ $(\delta_{\pi\pi} = 0.14 \pm 0.16$ \cite{pipi}).  Furthermore, the
ratio of $D \to KK$ isospin amplitudes 
$\left|\frac{\displaystyle A_1}{\displaystyle A_0}\right|$ is $3.5\sigma$ 
from one, which is the value obtained if only Standard Model spectator and 
penguin diagrams are allowed, exchange diagrams are neglected, and the FSI are 
entirely elastic.  Therefore, the substantial rate observed for 
$D^0 \to \bar{K^0} K^0$ can be attributed to one or more of
the following: (1) inelastic rescattering (FSI) from other modes
(e.g. $D^0 \to \phi\phi \to \bar{K}^0 K^0$ \cite{Kamal}) (2) large
contributions from W-exchange with $s\bar{s}$-popping in $D^0$ decay or 
(3) $D^+$ annihilation diagrams (unlikely in a factorization 
model \cite{BSW}). 

We gratefully acknowledge the efforts of the CESR staff in providing us with
excellent luminosity and running conditions.
This work was supported by 
the National Science Foundation,
the U.S. Department of Energy,
the Heisenberg Foundation,  
the Alexander von Humboldt Stiftung,
Research Corporation,
the Natural Sciences and Engineering Research Council of Canada,
and the A.P. Sloan Foundation.


\begin{references}
\bibitem{e9e25} The energy deposited by the photon candidate in a 3x3 block
of CsI crystals, normalized to the energy deposited in the 5x5 block
of crystals that surrounds the 3x3 block, must be greater than a 
value which is a function of shower energy.
\bibitem{norad}We use the result quoted in Ref. \cite{kmpp} 
that does not include radiative corrections.
\end{references}
\end{document}